\documentclass{aa}

\usepackage{graphicx}
\usepackage{natbib}
\usepackage{txfonts}

\newcommand{\lya}{Ly-$\alpha$}
\newcommand{\lyb}{Ly-$\beta$}

\begin{document}

  \title{Empirical relations between the intensities of Lyman lines of H and He$^+$.}
  \titlerunning{H and He$^+$ \lya\ lines correlations.}

  \author{M. Gordino\inst{1} \and F. Auch\`ere\inst{1} \and J.-C. Vial\inst{1} \and K. Bocchialini\inst{1} \and D. M. Hassler\inst{2} \and T.~Bando\inst{3} \and R. Ishikawa \inst{4} \and R. Kano\inst{3} \and K. Kobayashi\inst{5} \and N. Narukage\inst{3} \and J. Trujillo Bueno\inst{6} \and A. Winebarger\inst{5}}

  \offprints{F. Auch\`ere}

  \institute{Institut d'Astrophysique Spatiale,CNRS/Univ. Paris-Sud, Universit\' e Paris-Saclay, B\^at. 121, 91405 Orsay, France
   \and Southwest Research Institute, 1050 Walnut St., Suite 426, Boulder, CO 80302, USA
   \and National Astronomical Observatory of Japan, National Institutes of Natural Sciences, 2-21-1 Osawa, Mitaka, Tokyo 181-8588, Japan
   \and
   Institute of Space and Astronautical Science, Japan Aerospace Exploration Agency, 3-1-1 Yoshinodai, Chuo-ku, Sagamihara, Kanagawa 252-5210, Japan
   \and NASA Marshall Space Flight Center, ZP 13, Huntsville, AL 35812, USA
   \and Instituto de Astrofísica de Canarias, E-38205 La Laguna, Tenerife, Spain
   }

  \date{Received ; accepted}

  \abstract{Empirical relations between major UV and extreme UV spectral lines are one of the inputs for models of chromospheric and coronal spectral radiances and irradiances. They are also needed for the interpretation of some of the observations of the Solar Orbiter mission.}{We aim to determine an empirical relation between the intensities of the \ion{H}{I} 121.6~nm and \ion{He}{II} 30.4~nm \lya\ lines.}{Images at 121.6~nm from the Chromospheric Lyman-Alpha Spectro Polarimeter (CLASP) and Multiple XUV Imager (MXUVI) sounding rockets were co-registered with simultaneous images at 30.4~nm from the EIT and AIA orbital telescopes in order to derive a spatially resolved relationship between the intensities.}{We have obtained a relationship between the \ion{H}{I} 121.6~nm and \ion{He}{II} 30.4~nm intensities that is valid for a wide range of solar features, intensities, and activity levels. Additional SUMER data have allowed the derivation of another relation between the \ion{H}{I} 102.5~nm (\lyb) and \ion{He}{II} 30.4~nm lines for quiet-Sun regions. We combined these two relationships to obtain a \lya/\lyb\ intensity ratio that is comparable to the few previously published results.}{The relationship between the \ion{H}{I} 121.6~nm and \ion{He}{II} 30.4~nm lines is consistent with the one previously obtained using irradiance data. We have also observed that this relation is stable in time but that its accuracy depends on the spatial resolution of the observations. The derived \lya/\lyb\ intensity ratio is also compatible with previous results.}
  
  
  \maketitle

\section{Motivation\label{sec:motivations}}
The \lya\ and \lyb\ lines of neutral hydrogen respectively at 121.6~nm and 102.5~nm and the \lya\ line of singly ionized helium at 30.4~nm are among the brightest lines of the ultraviolet (UV) spectrum of the Sun, and their study is of importance in many areas of solar physics. These lines are observed by several instruments on board the Solar Orbiter mission \citep{Muller2020}. In this paper, we derive empirical relationships between these lines with, as described below, three main applications in mind: (i) the modeling of the resonance scattering emission of these lines in the corona, taking the nonuniformity of the chromospheric source  into account, (ii) the observational constraint of nonlocal thermodynamic equilibrium (NLTE) models of chromospheric structures, and (iii) the modeling of the UV and extreme UV (EUV) irradiances.

The bright coronal \lya\ line is mainly formed by the resonant scattering of the underlying chromospheric radiation by residual neutral hydrogen \citep{Gabriel1971}. The efficiency of the resonant scattering process depends on how much the chromospheric source profile and coronal scattering profile overlap. In a static atmosphere, the central wavelength of these two profiles lines up perfectly, whereas in a region with solar wind flow the scattering profile is Doppler-shifted relative to the disk profile, which results in a less efficient scattering and, therefore, in a reduced (dimmed) line intensity. Doppler dimming observations at \lya\ and in other lines that have a significant resonantly scattered component, such as the \ion{O}{vi}  doublet at 103.2 and 103.7 nm., have been extensively used to estimate the outflow velocities of the coronal plasma \citep[e.g.,][]{Kohl1997, Antonucci2000}. The interpretation of the \lya\ resonance scattering measurements relies on iterative forward modeling and thus requires independent knowledge of the chromospheric intensity, which is often taken from in-ecliptic irradiance measurements. However, \citet{Auchere2005b} has shown that using disk-integrated values and thus not taking into account the anisotropy of the illumination of the corona resulting from the distribution of bright features (e.g., active regions) and dark features (e.g., coronal holes) in the chromosphere can lead to significant systematic over- or underestimations of the coronal intensity, especially in polar regions. Recently, \cite{Dolei2018, Doleil2019} estimated that this translates to a $50\ \mathrm{km.s}^{-1}$ error on the outflow velocities derived from Doppler dimming methods. The same effect applies to all resonantly scattered coronal lines (i.e., the \lya\ line of \ion{He}{II} at 30.378~nm). Calibrated \lya\ disk images are thus in principle necessary for interpreting the coronal observations\footnote{This is also true of the modeling of the \lya\ radiation back-scattered by interplanetary neutral hydrogen \citep{Cook1981}.}. While 30.4~nm disk images are routinely available from narrowband telescopes such as the Extreme-ultraviolet Imaging Telescope \citep[EIT;][]{Delaboudiniere1995}, the Extreme Ultra Violet Imager \citep[EUVI;][]{Wuelser2004}, or the Atmospheric Imaging Assembly \citep[AIA;][]{Boerner2012, Lemen2012}, full-disk \lya\ observations are rare \citep[e.g.,][]{Bonnet1980}. \cite{Auchere2005b} thus introduced the possibility of using scaled 30.4~nm images as proxies for \lya\ data, which requires the derivation of an empirical relationship between the intensities of the two lines.

The chromospheric \lya\ line of hydrogen was first observed by \cite{Purcell1960} in a 1959 rocket flight and since then has been the subject of many more observations (see below), along with theoretical and modeling works. However, the profile of the line (very often self-reversed) has led to questions concerning both its characterization and interpretation. The line is optically thick and formed in non-local thermodynamic equilibrium \citep[NLTE;][]{Jefferies1961, Morton1961}, a situation that complicates the spectroscopic diagnostic. Nonetheless, the comparison of model outputs and observed line profiles of \lya , \lyb\ \citep[which is also self-reversed;][]{Tousey1963}, and other strong UV lines was at the root of one-dimensional (1D) models such as the Vernazza, Avrett, Loeser \citep[VAL;][]{Vernazza1981} and later the Fontenla, Avrett, Loeser~\citep[FAL;][]{Fontenla1990} models. Observations from the Orbiting Solar Observatory 8 (OSO 8) spacecraft and from sounding rockets \citep{Bonnet1978, Gouttebroze1978, Bonnet1981, Bonnet1982} of the \lya\ and \lyb\ lines have led to the modeling of the chromosphere, active regions \citep[e.g.,][]{Lemaire1981}, sunspots \citep{Kneer1981, Lites1982}, and prominences \citep{Vial1982}. In our present theoretical effort, the \lyb\ line was also used because it was simultaneously observed with the \lya\ line and provided a complementary diagnostic.
 
$\mathrm{H}^0$ and $\mathrm{He}^+$ have similar atomic structures, and NLTE modeling has shown that their \lya\ photons come from close-by but different layers of the solar atmosphere, although the details of the processes themselves may differ from one line to the other. \lya\ and \lyb\ lines of \ion{H}{I} are thermally produced (electronic collisions) in the chromosphere up to 30~000~K, and the \lya \ of \ion{He}{II} is formed above 30~000~K. On the contrary, in prominences, the \lya\ photons are mainly radiatively produced by resonance scattering. Radiation also plays an important role in the formation of the line of \ion{He}{II} since the existence of the $\mathrm{He}^+$ ion is determined by the incident EUV coronal radiation \citep[e.g.,][]{Andretta2003}. Moreover, the three lines play a very important role in the radiative losses of the various solar features \citep[see the VAL and FAL models for the 
chromosphere or the case of radiative equilibrium for prominences in ][]{Heinzel2012}. This means that simultaneous observations of the three abovementioned lines are (or would be) very useful for modeling the various solar structures. Despite the difficulties in recording the \lya\ line, nearly simultaneous observations of the \lya\ and \lyb\ lines of \ion{H}{I} were possible with the Solar Ultraviolet Measurement of Emitted Radiation \citep[SUMER;][]{Wilhelm1995} spectrograph, and~\cite{Lemaire2012} were able to constrain the ratio between the intensities of the two lines in various solar structures. A correlation between the intensities of the \lya\ lines of \ion{H}{I} and \ion{He}{II} has been obtained by \citet{Auchere2005b}, but, to the best of our knowledge, no relation has yet been established between the \lyb\ line of \ion{H}{I} and the \lya\ line of \ion{He}{II}.

Irradiance modeling is closely related to the two previous objectives. Semiempirical models of the solar spectral radiance adjust the variation in temperature with height in the solar atmosphere to obtain optimum agreement between calculated and observed continuum intensities, line intensities, and line profiles \citep{Avrett2008}. Major constraints to these models come from the observed intensities and profiles of the \lya\ and \lyb\ lines (and other strong UV lines) and in fact led to the development of the seminal 1D VAL and FAL models. For instance, the observations of the self-reversal of the \lyb\ line \citep{Tousey1963} constrained the models with the existence of a temperature plateau around 20 000~K.
Using this approach, the reconstruction of the spectral irradiance is obtained by a linear combination of a family of models in which the physical parameters are adjusted to match observations for a number of solar structures: coronal holes, quiet Sun, plage, sunspot umbra, penumbra, etc. It is thus necessary to use not only irradiance values, but also resolved measurements of the lines intensities.

In Sect.~\ref{sec:lya_hi_vs_heii} of the paper, we derive a new relationship between the intensities of the \lya\ lines of \ion{H}{I} and \ion{He}{II} using two different sets of spatially resolved data.  In Sect.~\ref{sec:lyb_hi_vs_heii} we derive a relationship between the intensities of the \lyb\ line of \ion{H}{I} and the \lya\ line of \ion{He}{II}. In Sect.~\ref{sec:lyb_hi_vs_lya_hi} we derive a relationship between the intensities of the \lyb\ and \lya\ lines of \ion{H}{I}. We summarize our findings in Sect.~\ref{sec:conclusions}.

\section{H\textsc{I} 121.6~nm versus He\textsc{II} 30.4~nm\label{sec:lya_hi_vs_heii}}

\cite{Auchere2005b} derived an empirical relation between the \ion{H}{I} 121.6~nm and \ion{He}{II} 30.4~nm irradiances using data from the EIT telescope, the Solar Stellar Irradiance Comparison Experiment \citep[SOLSTICE;][]{Rottman1993}, and the Solar EUV Experiment \citep[SEE;][]{Woods2000}. It matched spatially resolved but episodic observations from Skylab \citep{Vernazza1978}. In the present study, we revisit this relationship using two new sets of spatially resolved observations that cover all types of solar structures and different activity levels. The first set is full-disk observations made on November 2, 1998, by EIT and the Multiple XUV Imager \citep[MXUVI;][]{Auchere1999}. The second set corresponds to disk-center and limb observations made on September 3, 2015, by the AIA and the Chromospheric Lyman-Alpha Spectro Polarimeter \citep[CLASP;][]{Kano2012, Kobayashi2012, Kano2017}.

\subsection{MXUVI and EIT data}

\begin{figure*}
\centering
\includegraphics[width=\textwidth]{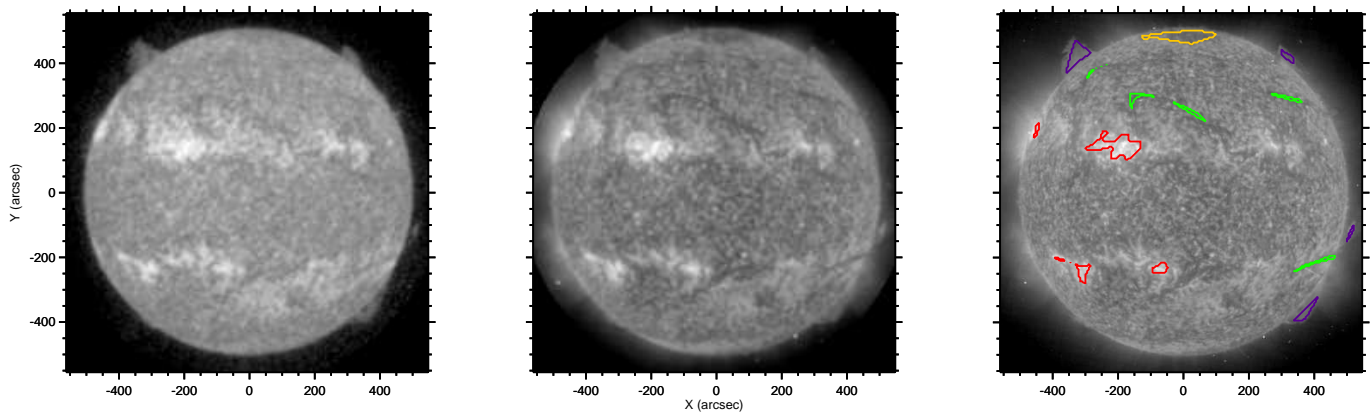}
\caption{Quasi-simultaneous images of the Sun taken on November 2, 1998, in \ion{H}{I} 121.6~nm (left) and \ion{He}{II} 30.4~nm (center, right). The Lyman-$\alpha$ image was obtained by the rocket-borne MXUVI instrument at 18:20 UT. The 30.4~nm image was recorded by EIT at 18:26 UT. The EIT image was degraded to match the resolution of the MXUVI. Solar north is up. The intensity scales are logarithmic in both cases, and the contrast is the same. There is a remarkable morphological similarity for all types of structures, but, as is also visible in Fig.~\ref{fig:eit-mxuvi-correl}, they are less contrasted at Lyman~$\alpha$. In the right image, we overlaid different contours corresponding to different solar regions: active regions (red), filaments (green), the coronal hole (yellow), and prominences (purple).}
\label{eit-mxuvi}
\end{figure*}

The second flight of the MXUVI sounding-rocket-borne telescope occurred on November 2, 1998, at 18:20 UT. The \lya\ channel of the MXUVI uses a 256 $\times$ 256 detector with a 10 arcsec/pixel sampling. Isolation of the \lya\ line is obtained using Al/MgF2 filters and selective mirror coating for a resulting passband of 10~nm. All the data acquired during the flight were co-registered and added into a single image. The flat field and dark level were corrected according to \citet{Auchere1999}. Radiometric calibration was obtained by scaling the resulting image to the simultaneous composite \lya\ irradiance, which for this date includes data from SOLSTICE \citep{Rottman1993, Woods1993}. The composite passband has a width of 1.0~nm.

The EIT telescope has been providing Sun full-disk observations since 1996. We used the November 2, 1998, image taken at 18:26 UT, which is the 30.4~nm image closest in time to the MXUVI second flight and corresponds to a specific EIT campaign for this flight. The 30.4~nm channel uses a 1024 $\times$ 1024 detector with a 2.627 arcsec/pixel sampling \citep{Auchere2000a, Auchere2004}. The flat field, dark noise, and degradation were corrected with the \texttt{eit\_prep} procedure. Radiometric calibration was obtained by using the \texttt{eit\_parms} function. The instrument passband has a width of 7.3~nm.

\subsection{CLASP and AIA data}

The first flight of the CLASP sounding-rocket-borne spectro-polarimeter occurred on September 3, 2015, between 17:03 UT and 17:08 UT. The \lya\ images used are provided by the CLASP Slit-Jaw camera (CLASP-SJ). The CLASP-SJ uses a 528 $\times$ 536 pixel sensor with a 1.03 arcsec/pixel sampling. The  instrument passband has a width of 3.5~nm. The flat field and dark level were corrected. The radiometric calibration was obtained by scaling the resulting image to the simultaneous composite \lya\ irradiance, which for this date includes data from SOLSTICE\citep{McClintock2005a, McClintock2005b, Snow2005}. The composite passband has a width of 1.0~nm. During this flight, images from the Sun center and limb were taken.

The AIA telescope has been providing full-disk observations at high resolution since 2010. The 30.4~nm channel of AIA uses a 4096 $\times$ 4096 detector with a 0.63 arcsec/pixel sampling. The 12-second cadence permitted 25 AIA images to be obtained during the CLASP flight. The radiometric calibration was obtained by using the \texttt{aia\_get\_response} function. The instrument passband has a width of 4.5~nm.

\subsection{Data co-registration\label{sec:co-registration}}

\begin{figure*}
\centering
\includegraphics[width=\textwidth]{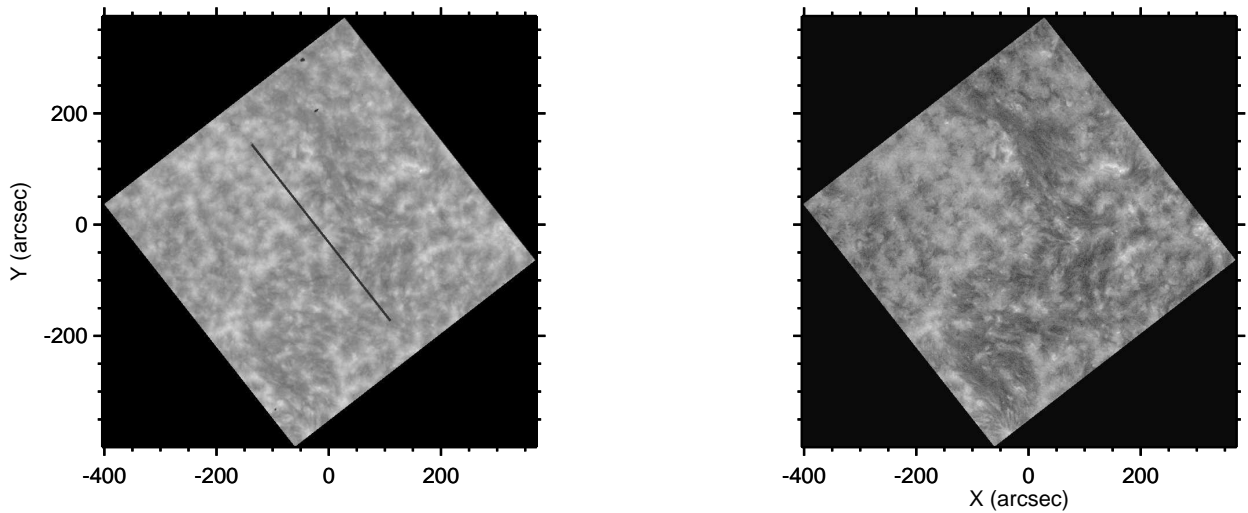}
\caption{Sun center \ion{H}{I} 121.6~nm observations from September 3, 2015, from the CLASP-SJ (left) and simultaneous \ion{He}{II} 30.4~nm observations from AIA degraded to match the CLASP-SJ resolution (right). The black band in the center of the CLASP-SJ image is the entrance slit of the spectrometer. Solar north is up.}
\label{aia-clasp-obs-cen}
\end{figure*}

\begin{figure*}
\centering
\includegraphics[width=\textwidth]{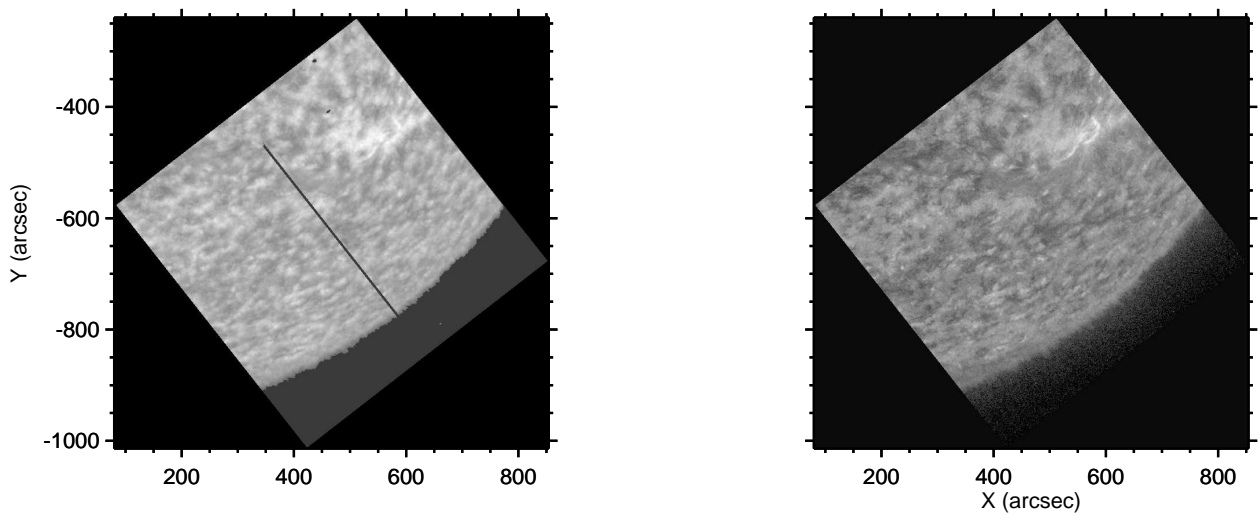}
\caption{Same as Fig.~\ref{aia-clasp-obs-cen},  but for the CLASP-SJ limb pointing.}
\label{aia-clasp-obs-limb}
\end{figure*}

We co-aligned the \ion{He}{II} 30.4~nm (EIT or AIA) and \ion{H}{I} 121.6~nm (MXUVI or CLASP-SJ) images to compare and find a relation between the intensities. We used a reduced $\chi^{2}$ cross-correlation method in translation and rotation. The spatial scaling was assumed to be known from the independently derived plate scales and from the distances between the instruments and the Sun. When scanning the parameter space, the highest resolution images (EIT and AIA) were resampled using bilinear interpolation in the reference frame of the lowest resolution images (MXUVI and CLASP), the latter being left untouched. For CLASP-SJ, each image was co-registered with the AIA image that was closest in time. We validated our procedure by co-aligning an EIT 30.4~nm image with an AIA 30.4~nm image taken on August 1, 2011, at 01:07 UT. Since the pointing information of the two instruments is known independently from the headers of the images, this test shows that the co-alignment error is about 0.15 pixels rms. This translates to 0.16 arcsec for AIA-CLASP pairs and 1.6 arcsec for EIT-MXUVI pairs. These sub-pixel co-alignment errors contribute negligible dispersion to the correlations described in Sect.~\ref{sec:heii_lya_results}. Figure~\ref{eit-mxuvi} shows the \ion{H}{I} 121.6~nm MXUVI and the \ion{He}{II} 30.4~nm EIT images after co-registration. The first two images were used to find the relationship between the intensities of the two lines and correspond to MXUVI and EIT images with a 10 arcsec pixel resolution. In the third image, which shows the original full resolution EIT data, colored contours correspond to different solar regions. Figures~\ref{aia-clasp-obs-cen} and \ref{aia-clasp-obs-limb} show the \ion{H}{I} 121.6~nm CLASP-SJ and the \ion{He}{II} 30.4~nm AIA images after co-registration for, respectively, Sun center and limb observations.

\subsection{Isolation of the He\textsc{II} 30.4~nm line\label{sec:heii_isolation}}

In addition to the main line of \ion{He}{II} at 30.378~nm, the 30.4~nm passbands of EIT and AIA include several coronal and transition region lines. In particular, the \ion{Si}{XI} line at 30.332~nm cannot be suppressed by the multilayer coating technology of these instruments. In order to estimate the contribution of all the contaminating spectral lines from the EIT and AIA, we constructed a differential emission measure (DEM) curve for each pixel of the images. Then, knowing the spectral response of the instruments, we computed the intensity of all the spectral lines included in the passbands except for that of \ion{He}{II} at 30.378~nm. The sum of their contributions was then removed from the original 30.4~nm image to obtain the intensity of the \ion{He}{II} 30.378~nm line alone. For EIT, the DEM curve was constructed using the code developed specifically for EIT by \citet{Cook1999} and used in \citet{Auchere2005} to compute the \ion{He}{II} irradiance. For AIA, we used the Gaussian DEM inversion from \citet{Guennou2012a, Guennou2012b}. As a validation, we verified that the computed intensity of the \ion{Si}{XI} line represents 5 to 20\% of that of the \ion{He}{II} line, in agreement with the spectroscopically derived values of \citet{Thompson2000}.

\subsection{Results\label{sec:heii_lya_results}}
\begin{figure*}
\centering
\includegraphics[width=\textwidth]{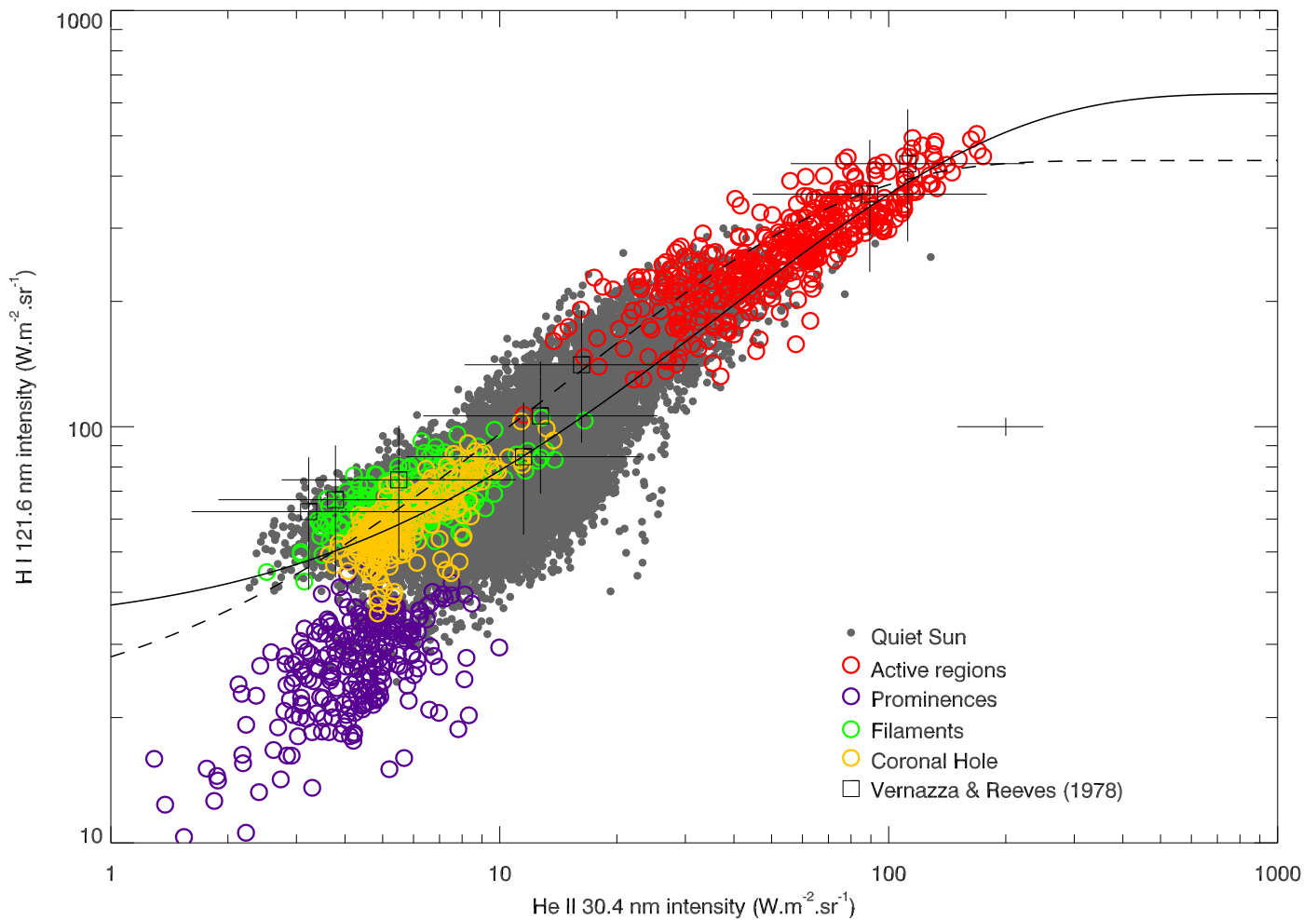}
\caption{Correlation between \ion{He}{II} 30.4~nm and \ion{H}{I} 121.6~nm for the November 2, 1998, data set (MXUVI and EIT). The cross on the right indicates the error bars for each data point. The different colored circles on the plot correspond to the intensities of the colored contours in the EIT image (Fig. \ref{eit-mxuvi}). The dark squares are the Vernazza \& Reeves (1978) data from Skylab, shown with their error bars. The solid line fits the MXUVI and EIT data. The dashed line corresponds to the fit from \cite{Auchere2005b}.}
\label{fig:eit-mxuvi-correl}
\end{figure*}

In Fig.~\ref{fig:eit-mxuvi-correl} we have represented the intensity of the \ion{H}{I} 121.6~nm line measured by MXUVI as a function of that of the \ion{He}{II} line at 30.4~nm measured by EIT. We fitted the data points with the following function:

\begin{equation}
         I_{121.6} = C_{1}(1-C_{2}e^{-C_{3}I_{30.4}}) \ [\mathrm{W}.\mathrm{m}^{-2}.\mathrm{sr}^{-1}]
\label{eq1}
.\end{equation}

\begin{figure*}
\centering
\includegraphics[width=\textwidth ]{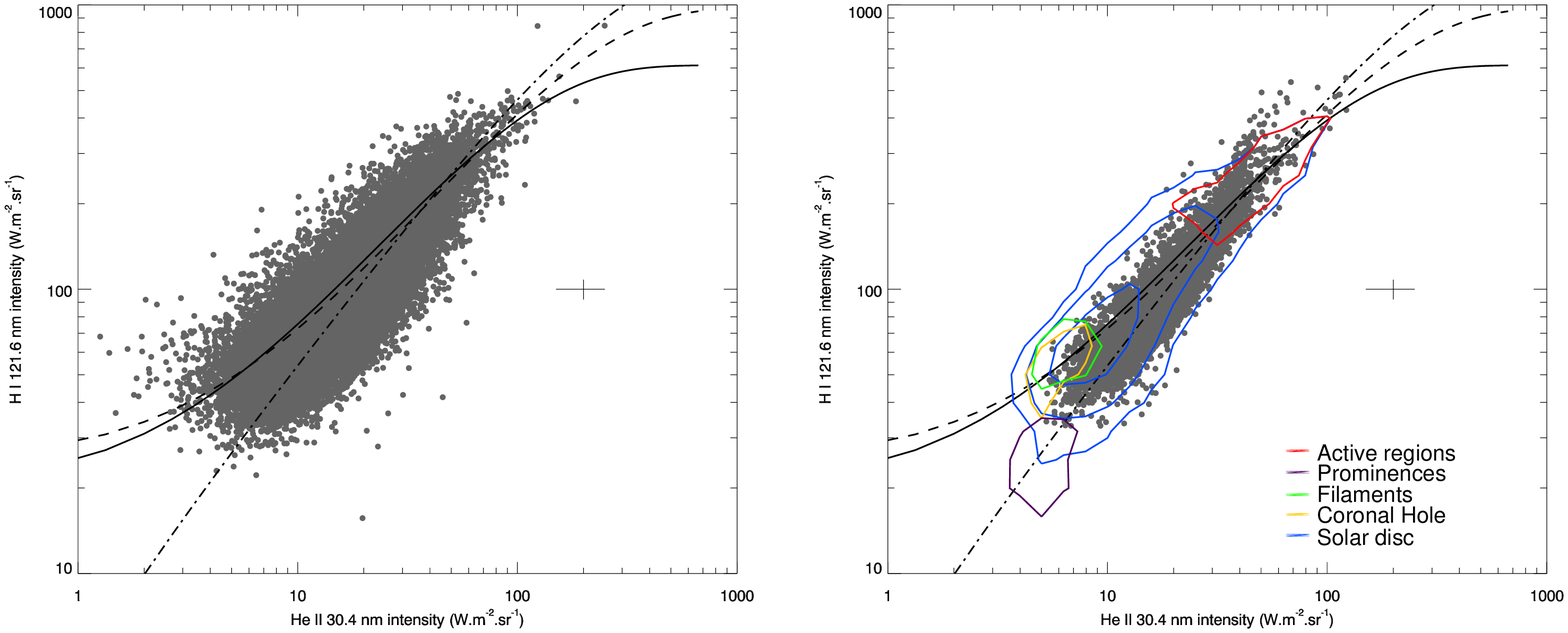}
\caption{Correlation between \ion{He}{II} 30.4~nm and \ion{H}{I} 121.6~nm for the September 3, 2015, data set (CLASP-SJ and AIA). The plot in the left panel corresponds to the 1.03 arcsec/pixel CLASP-SJ resolution, whereas the plot in the right panel has the 10 arcsec/pixel MXUVI resolution. The cross on the right of each plot indicates the error bars for each data point. The solid lines correspond to the fit shown in Fig.~\ref{fig:eit-mxuvi-correl}. The dot-dashed lines correspond to the fit of the CLASP (at MXUVI resolution) and AIA data. The dashed lines show the fit for both the MXUVI-EIT and CLASP-AIA data. The colored contours outline the intensity distributions for the different solar regions from Fig.~\ref{fig:eit-mxuvi-correl}.}
\label{aia-clasp-correl}
\end{figure*}

This relation was introduced by \cite{Vourlidas2001}, and it was used by \cite{Auchere2005b} to fit the correlation between the \ion{H}{I} 121.6~nm and \ion{He}{II} 30.4~nm irradiances. The best-fit parameters are given in Table~\ref{Coefs} along with those obtained by \cite{Auchere2005b}. The solid and dashed curves in Fig.~\ref{fig:eit-mxuvi-correl} correspond respectively to the present fit and to that of \cite{Auchere2005b}. The colored circles correspond to the contribution of the different solar regions encircled in the third image of Fig.~\ref{eit-mxuvi}. As observed by \cite{Auchere2005b}, we notice in active regions (red circles) a saturation effect at high intensities. The prominence data points have a distinct distribution because of the predominantly radiative formation process of that line compared to other types of structures. It should be noted that the data points corresponding to prominences were not taken into account for the fit. Indeed, for the primary application of coronal modeling, prominences represent a negligible contribution. The uncertainty for EIT intensities is about 25\%, and the uncertainty for the \lya\ irradiance from SOLSTICE is about 5\% for November 2, 1998 \citep{Snow2016}. The offset between the two fits, at most 30\%, can thus be explained by calibration uncertainties. We also plot the data from Skylab \citep{Vernazza1978} (squares), which match with the two relations found. This shows that the correlation between \ion{He}{II} 30.4~nm intensity and \ion{H}{I} 121.6~nm intensity is stable with time.\\

\begin{table}
\centering
\begin{tabular}{l|l|l|l}
& $C_{1}$ & $C_{2}$ & $C_{3}$\\
\hline  MXUVI - EIT & 613 & 0.968 & 0.00991\\
\hline  CLASP - AIA & 1461 & 1.002 & 0.00422\\
\hline  Both data sets & 820 & 0.975 & 0.00681\\
\hline \cite{Auchere2005b} & 436 & 0.955 & 0.0203\\
\end{tabular}
\caption{Comparison between the fitted coefficients of Eq.~\ref{eq1} for the data sets considered in the present work and those in \cite{Auchere2005b}.}
\label{Coefs}
\end{table}

Figure~\ref{aia-clasp-correl} shows the intensity of CLASP-SJ \ion{H}{I} 121.6~nm as a function of AIA \ion{He}{II} 30.4~nm intensity. The uncertainties for AIA 30.4~nm are about 25\% \citep{Boerner2012}. Regarding the \lya\ irradiance from SOLSTICE, the uncertainties are about 8\% for September 3, 2015 \citep{Snow2016}. The two panels correspond to two spatial resolutions: the CLASP-SJ resolution (1.03~arcsec, left) and binned to the MXUVI resolution (10~arcsec, right). The solid lines in the two plots correspond to the relation derived in the previous section from MXUVI and EIT data. The dotted-dashed line corresponds to the relation derived from CLASP and AIA images, and the dashed line corresponds to the relation derived from both data sets. These fits have been calculated with images at the MXUVI resolution. The colored contours correspond to the clusters of colored circles in Fig.~\ref{fig:eit-mxuvi-correl}. They show that the relations we obtained from CLASP-SJ and AIA images can also describe the correlation between the \ion{H}{I} 121.6~nm and \ion{He}{II} 30.4~nm lines from MXUVI-EIT images. This confirms, as mentioned above, that the relation between \ion{He}{II} 30.4~nm and \ion{H}{I} 121.6~nm intensities is stable in time.

The best-fit relationships of Figs.~\ref{fig:eit-mxuvi-correl} and \ref{aia-clasp-correl} are consistent given the uncertainties of each data set. While being thus equivalent, the corresponding uncertainties must be propagated when being used for the possible applications listed in Sect.~\ref{sec:motivations}. We also notice that there is more dispersion in the left plot of Fig.~\ref{aia-clasp-correl}, which corresponds to the  higher resolution images. It shows that the correlation is stronger when one observes at lower spatial resolution. This is consistent with the extreme case of the disk-integrated data used by \cite{Auchere2005b}, for which the dispersion was even smaller. Consequently, the empirical relationships given here are particularly suited for applications that do not require high spatial resolution, such as the coronal and irradiance modeling described in Sect.~\ref{sec:motivations}. The increase in dispersion with increased resolution is possibly a signature of the differences between the formation processes of the two lines. We also notice that the dispersion is larger in \ion{He}{II} than in \ion{H}{I}. This could be explained by the sensitivity of the Planckian (collisional) contribution to the source function, since its sensitivity to temperature is proportional to the frequency. There is also a clearly different behavior between the quiet Sun and prominences for a given \ion{He}{II} intensity (see Fig.~\ref{fig:eit-mxuvi-correl}). The study of these differences is potentially important for chromospheric modeling but is beyond the scope of this paper.

\section{H\textsc{I} 102.5~nm versus He\textsc{II} 30.4~nm\label{sec:lyb_hi_vs_heii}}

\subsection{SUMER and EIT data}

In order to derive this relationship, we used data from the SUMER instrument. We used a \lyb\ line raster measured on April 8, 1996, between 06:09 UT and 06:36 UT, close to the Sun center. The spectra we used were recorded by the SUMER/A detector, which has 1024 (spectral) $\times$ 360 (spatial) pixels$^{2}$. The slit used for the raster was the 1 $\times$ 120 arcsec$^{2}$ one. The raster contains 313 spectra at the \lyb\ line with a step of 0.38~arcsec between each spectrum. Each spectrum covers a 0.7~nm width. We used the standard preparation routine, \texttt{sum\_read\_corr\_fits}, which includes the radiometric calibration. For the 30.4~nm data, we used the April 8, 1996, EIT image taken at 06:21 UT, which is the image closest in time to the SUMER raster.

\begin{figure*}
\centering
\includegraphics[width=\textwidth]{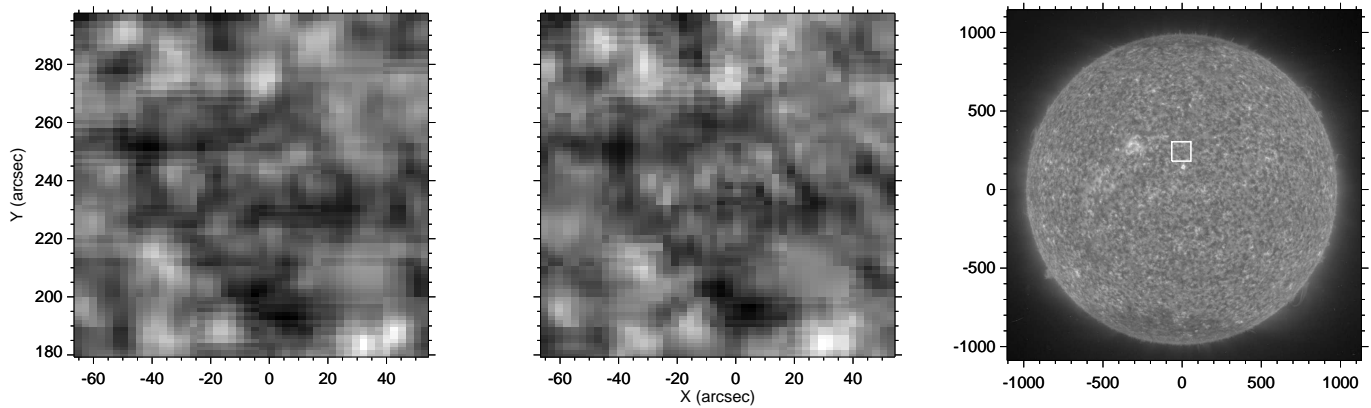}
\caption{Observations from SUMER at \ion{the H}{I} Lyman-$\beta$ line on April 8, 1996, at 06:09:27 UT (left) and EIT at \ion{the He}{II} 30.4~nm line on April 8, 1996, at 06:21:51 UT, corresponding to the SUMER field of view (middle) and the full-disk observation (right). The white square on the full-disk EIT image shows the position of the SUMER field of view.}
\label{eit-lb obs}
\end{figure*}

\subsection{Data processing}

In order to build the SUMER raster image, we integrated the intensity along the wavelength axis to obtain the total intensity at each point of the field of view. We then degraded the resulting SUMER image by smoothing it down to the EIT 2.627~arcsec resolution. We used the same co-alignment procedure used for the study of the correlation between \ion{He}{II} 30.4~nm and \ion{H}{I} 121.6~nm intensities (Sect.~\ref{sec:co-registration}). Figure \ref{eit-lb obs} shows the resulting \ion{H}{I} 102.5~nm SUMER image (left) along with the co-registered EIT image (middle). These two images were used to find the relationship between the two lines. The third image (right) shows the original EIT data, the white square indicating the region covered by the SUMER raster.

\subsection{Results}

\begin{figure}
\centering
\includegraphics[width=0.5\textwidth]{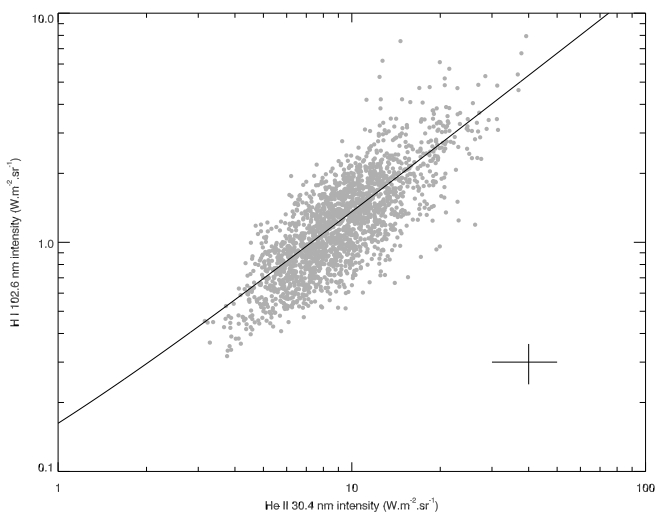}
\caption{Correlation between \ion{He}{II} 30.4~nm from EIT and \ion{H}{I} 102.5~nm from SUMER. The cross at the bottom right indicates the error bars for each data point. The solid line is an estimation of the correlation with a linear fit.}
\label{eit-lb correl}
\end{figure}

Figure \ref{eit-lb correl} shows the SUMER \ion{H}{I} 102.5~nm  intensity as a function of the EIT \ion{He}{II} 30.4~nm intensity. The solid curve corresponds to a linear fit to the data with the following relation:
\begin{equation}
        I_{102.5} =  0.132 I_{30.4} + 0.0233 \ [\mathrm{W}.\mathrm{m}^{-2}.\mathrm{sr}^{-1}]
\label{eq2}
.\end{equation}

The uncertainties for SUMER intensities are about 20\%. Equation~\ref{eq2} was obtained using only quiet-Sun data, covering a limited range of intensities, which justifies the linear fit. We note that the ratio ${I_{30.4}}/{I_{102.5}}$ is about 8 in the quiet Sun, to be compared to the irradiance ratio \citep[in the range 5-10, e.g., in June 2010;][]{ssisee}, the \cite{Vernazza1978} ratio of 11, or the X-flare ratio \citep[about 5;][]{Milligan2012}.

\section{H\textsc{I} 102.5~nm versus H\textsc{I} 121.6~nm\label{sec:lyb_hi_vs_lya_hi}}

\begin{figure*}
\centering
\includegraphics[width=0.9\textwidth]{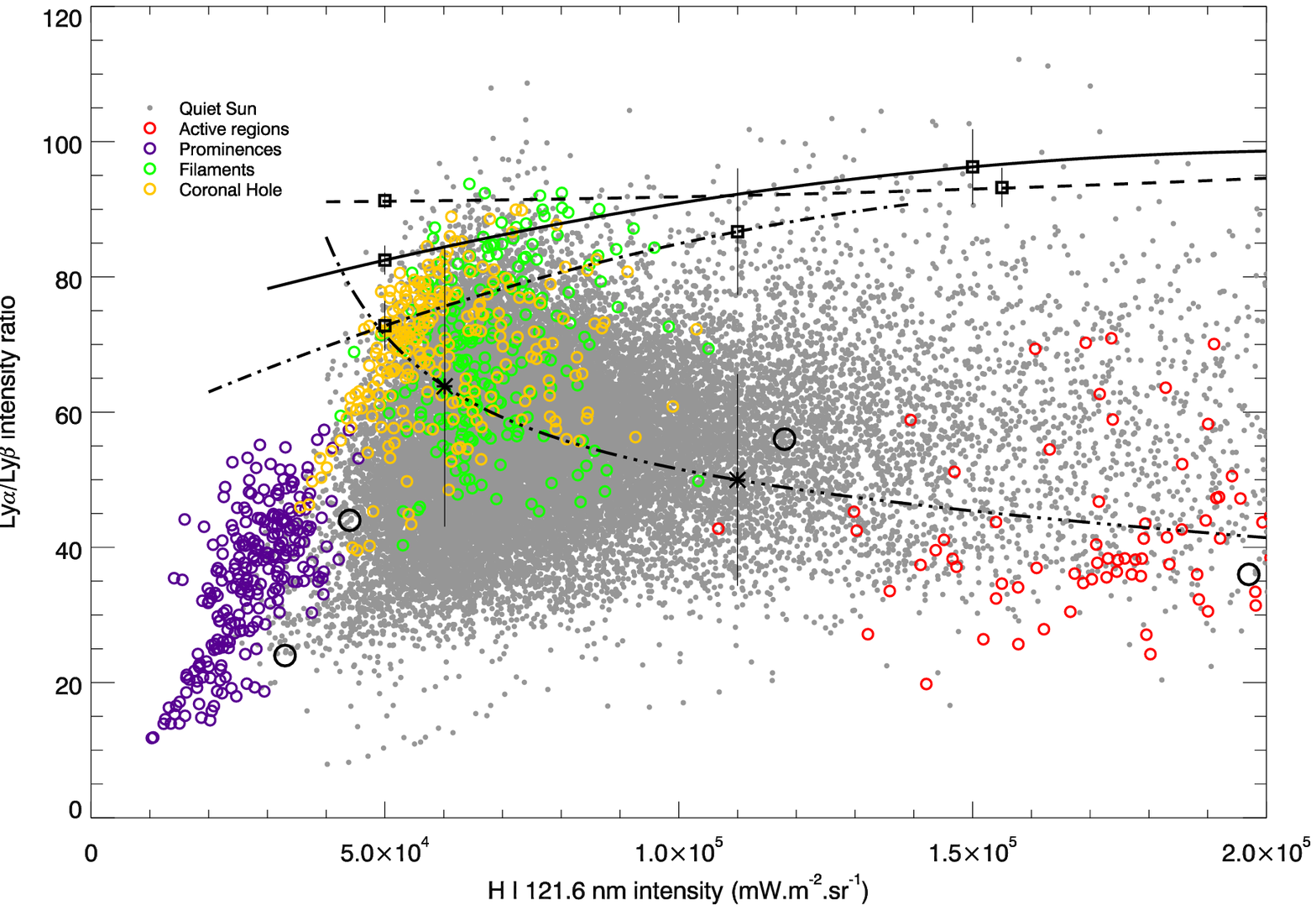}
\caption{Distribution of the \ion{H}{I} \lya/\lyb\ intensity ratio from the MXUVI image and the EIT image converted to \lyb\ using Eq. ~\ref{eq2}. The dots and colored circles represent the measurements, which were performed at a period of minimum activity. The colored circles correspond to the different solar regions defined in Fig.~\ref{fig:eit-mxuvi-correl}. In order to maintain SI units, \lya\ intensities are expressed in $\mathrm{mW}.\mathrm{m}^{-2}.\mathrm{sr}^{-1}$, providing numerical values equivalent to the centimeter-gram-second units of $\mathrm{erg}.\mathrm{s}^{-1}.\mathrm{cm}^{-2}.\mathrm{sr}^{-1}$ used by \cite{Lemaire2012}. The three-dotted-dashed line represents the \ion{H}{I} \lya /\lyb\ intensity ratio deduced from Eqs.~\ref{eq1} and \ref{eq2}. The stars with error bars give an estimate of the accuracy of the ratio. The three other lines are the fits from Fig. 2 of \cite{Lemaire2012}. The solid and dashed lines correspond to two quiet-Sun observations, whereas the dotted-dashed line corresponds to a coronal hole. The squares with error bars on these lines give an estimation of the error for different ratio values on the three fits. The large black circles correspond to the ratio deduced from \cite{Patsourakos1998}.}
\label{la-lb ratio}
\end{figure*}

\cite{Lemaire2012} measured the \lya /\lyb\ ratio from SUMER data, which were obtained in coronal holes, the quiet Sun, and active regions. As we have two relationships, between \ion{He}{II} and \ion{H}{I} \lya\, and between \ion{He}{II} and \ion{H}{I} \lyb, we can indirectly obtain a relationship between \ion{H}{I} \lya\ and \ion{H}{I} \lyb. 
In order to do this, we used the November 2, 1998, EIT 30.4~nm image converted to \lyb\ using Eq.~\ref{eq2}. This requires the assumption that Eq.~\ref{eq2}, which was derived for April 8, 1996, still holds for November 2, 1998. We also suppose that Eq.~\ref{eq2}, obtained for the quiet Sun, can be extrapolated to regions of higher intensities.
In order to compare our results to the ones of \cite{Lemaire2012}, we plot in Fig.~\ref{la-lb ratio} the \lya/\lyb\ intensity ratio as a function of the \lya\ intensity. We chose a range that matches the range of validity of Eq. \ref{eq2}. Using Eqs. \ref{eq1} and \ref{eq2}, we can derive a relation for the \lya/\lyb\ intensity ratio. This relation is represented in Fig. \ref{la-lb ratio} by the three-dotted-dashed line. As in Sect.~\ref{sec:lya_hi_vs_heii}, prominences are not taken into consideration to derive Eq.~\ref{eq1}. We trace the three fits from \cite{Lemaire2012}, which correspond to one coronal hole and two quiet-Sun observations. We notice higher ratios for coronal holes and lower ratios for active regions than for the quiet Sun. We also plot values that can be derived from the intensities obtained by \citet{Patsourakos1998} from OSO 8 spectra. These values are represented by the large circles and correspond respectively to, from left to right, cell, network, bright point, and plage. This shows that our \lya/\lyb\ ratio is in the range of values measured by \cite{Lemaire2012} and deduced from \cite{Patsourakos1998}. However, they do not agree with the higher values derived by \cite{Tian2009b} and \cite{Tian2009a} for the quiet Sun (about 200) and a coronal hole (130). An explanation of these discrepancies is given in \cite{Lemaire2012}. Given the uncertainties of about 50\% on the \lya/\lyb\ ratio, represented by the stars with error bars, our values for coronal holes are compatible with those of \cite{Lemaire2012}. However, as pointed out by \cite{Bocchialini1994, Bocchialini1996} and \cite{Patsourakos1997}, the \lya\ and \lyb\ lines appear to have a singular behavior in equatorial coronal holes. For active regions, the ratio is lower. This can be explained if the linear relationship between \ion{H}{I} 102.5~nm and \ion{He}{II} 30.4~nm is not valid for active regions. To be fully compatible with \cite{Lemaire2012}, including for active regions, we can predict that the relationship between \ion{He}{II} 30.4~nm and \ion{H}{I} 102.5~nm in fact saturates at high intensities, such as the one found between \ion{He}{II} 30.4~nm and \ion{H}{I} 121.6~nm (Sect.~\ref{sec:lya_hi_vs_heii}). We note that the \lya/\lyb\ ratio is much higher (210) in a streamer at about 1~R$_{\odot}$ above the surface \citep{Giordano2013}. Such a high value could be explained by the importance of resonance scattering in both lines. We finally mention that in the case of a coronal mass ejection (CME) found a ratio of about 4, while \cite{Ciaravella2003} report a ratio of 450 in the pre-CME and CME phases.

\section{Conclusions\label{sec:conclusions}}

Using full-disk images, we derived an empirical relationship between \ion{H}{I} \lya\ and \ion{He}{II} 30.4~nm intensities. Considering the uncertainties and the scatter on the intensities, our results are compatible with those obtained by \cite{Auchere2005b} from irradiance measurements, as well as with the spatially resolved results of \cite{Vernazza1978}. As the observations span four decades and various levels of activity, we conclude that this relationship is stable in time. We also observed that the dispersion around the average relationship increases with increasing resolution, which could have implications for NLTE modeling. We derived a new relation between the \ion{H}{I} \lyb\ and \ion{He}{II} 30.4~nm intensities for two quiet-Sun  regions. Combined with the \ion{H}{I} \lya\ and \ion{He}{II} 30.4~nm relationship, we also obtained a \lya/\lyb\ intensity ratio compatible with previous results in the quiet Sun, which can be used to constrain chromospheric emission models. Our results are not only important for the modeling of various solar features, but also for the reconstruction of the spectral irradiance in \ion{H}{I} and \ion{He}{II} lines, which are important radiative input components in the Earth's ionosphere and thermosphere.

The three lines studied in this paper can be observed simultaneously by the Extreme Ultraviolet Imager~\citep[EUI;][]{Halain2015, Rochus2020} telescope, the Spectral Imaging of the Coronal Environment~\citep[SPICE;][]{Spice2020} spectrometer, and the Metis \citep{Antonucci2012, Antonucci2020} coronagraph on Solar Orbiter~\citep{Muller2013a, Muller2020}. The derived relationships will help in the joint analysis of the observations of these three instruments~\citep{Auchere2020}. In particular, the relationships derived in this paper are needed to analyze the Metis observations of the resonantly scattered emission of \ion{H}{I} in the corona. Since Metis does not measure the \lya\ disk intensity, it is possible to instead use 30.4~nm data from the full-Sun channel of the EUI telescope \citep{Auchere2005c} to compute the illumination of the corona at 121.6~nm.

\bibliographystyle{aa}

\bibliography{bibliography}

\end{document}